\documentclass[pre,a4paper,amsfonts,amssymb,onecolumn,showpacs,showkeys,nofootinbib,floats]{revtex4}

\usepackage{amsfonts}
\usepackage{amsmath}
\usepackage{amssymb}
\usepackage{graphicx}

\begin{document}

\title{On the multi-fractal structure of traded volume in financial markets}

\author{L.~G.~Moyano}
\email[e-mail address: ]{moyano@cbpf.br}
\affiliation{Centro Brasileiro de Pesquisas F\'{\i}sicas, 150, 22290-180, Rio de Janeiro - RJ, Brazil}
\author{J.~de~Souza}
\email[e-mail address: ]{jeferson@cbpf.br}
\affiliation{Centro Brasileiro de Pesquisas F\'{\i}sicas, 150, 22290-180, Rio de Janeiro - RJ, Brazil}
\author{S.~M.~Duarte~Queir\'{o}s}
\email[e-mail address (Corresponding author): ]{sdqueiro@cbpf.br}
\affiliation{Centro Brasileiro de Pesquisas F\'{\i}sicas, 150, 22290-180, Rio de Janeiro - RJ, Brazil}

\date{\today}

\begin{abstract}
In this article we explore the multi-fractal properties of 1 minute traded volume of the 
equities which compose the Dow Jones 30. We also evaluate the weights of 
linear and non-linear dependences in the multi-fractal structure of the observable.
Our results show that the multi-fractal nature of traded volume comes essencially from 
the non-Gaussian form of the probability density functions and from non-linear dependences.
\end{abstract}



\maketitle

\section{Introduction}

The intricate character of financial markets has been one of the main
motives for the physicists interest in the study of their statistical and
dynamical properties~\cite{voit,gm-ct}. Besides the asymptotic power-law
behaviour for probability density function for price fluctuations, \textit{%
the return}, and the long-lasting correlations in the absolute return,
another important statistical feature observed is the return multi-fractal nature~\cite{mandelbrot}. 
This propriety has been important in the
establishment of analogies between price fluctuations and fluid turbulence~\cite{gashghaie} 
and the development of multiplicative cascade models for
return dynamics too~\cite{arneodo}.

Changes in the price of a certain equity are basically related transactions
of that equity. So, the \textit{traded volume}, which is defined as the
number of stocks that change hands during some period of time, is an
important observable in the dynamics of financial markets. This observation
is confirmed by an old proverb at Old Street that "It takes volume to make
prices move"~\cite{karpoff}.

In previous works several proprieties of traded volume, $V$, either
statistical or dynamical have been studied~\cite{volume,vol-smdq,vol-dumas}.
In this article, we present a study of the multi-fractal structure of $1$%
-minute traded volume time series of the $30$ equities which are used to
compose the Dow Jones Industrial Average, DJ30. Our series run from the $%
1^{st}$ of July until the $31^{st}$ December of $2004$ with a length of
around $50k$ elements each. The analysis is done using the Multi-Fractal
Detrended Fluctuation Analysis, MF-DFA
\footnote{To evaluate the local trends 
we have used $5^{th}$-order polynomials. From this order ahead, we have a nearly
polynomial-order-independent multi-fractal spectrum, contrarly to what 
happens with fitting polynomials of smaller order.}
~\cite{mf-dfa}. Besides the
multi-fractral analysis we weight the influence of correlation, asymptotic
power-law distribution and non-linearities in the multi-fractality of traded
volume. Since we are dealing with intra-day series we have to be cautious
with the well-known daily pattern which is often considered as a lacklustre
propriety\cite{admati}. To that, we have removed that intra-day pattern of
the original time series and normalised each element of the series by its
mean value defining the normalised traded volume, $v\left( t\right) =\frac{%
V^{\prime }\left( t\right) }{\left\langle V^{\prime }\left( t\right)
\right\rangle ,}$ where $V^{\prime }\left( t\right) =\frac{V\left( t\right) 
}{\Xi \left( t^{\prime }\right) }$, $\Xi \left( t^{\prime }\right) =\frac{%
\sum\limits_{i=1}^{N}V\left( t_{i}^{\prime }\right) }{N}$ and $\langle
\ldots \rangle $ is defined as the average over time ($t^{\prime }$
represents the intra-day time and $i$ the day).
\section{Multi-fractality and its components}
\begin{figure}[tbp]
\begin{center}
\includegraphics[width=0.95\columnwidth,angle=0]{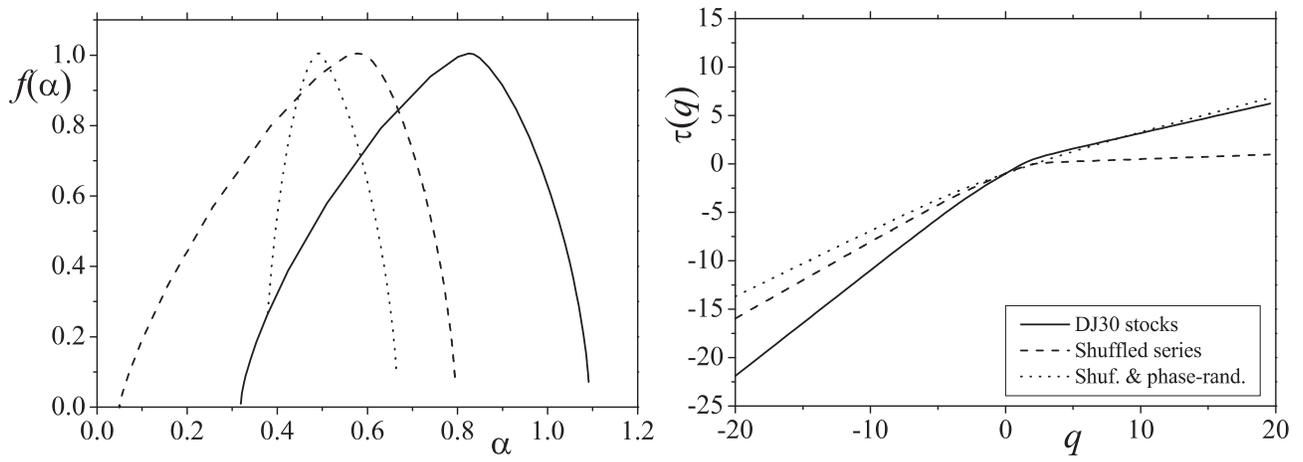}
\end{center}
\caption{Left panel: Multi-fractal spectra $f\left( \protect\alpha \right) $
\textit{vs. }$\protect\alpha $. Right panel: Scaling exponents $\protect%
\tau \left( q\right) $. \textit{vs. }$q$ averaged over the $30$ companies.
The legend on the right is also valid for the left. The "original" and
shuffled time series present a strong multi-fractal character, while the
shuffled plus phase randomised time series presents a narrow width in $\alpha $ related
with the almost linear dependence of $\protect\tau _{sh-ran}\left( q\right) $
and also with the strong contribution of non-Gaussian PDF of traded volume   
to multi-fractality.} \label{fig}
\end{figure}

A common signature of complexity in a system is the existence of
(asymptotic) scale invariance in several typical quantities. 
This scale invariance, self-affinity for time series, can be
associated to a single type of structure, characterised by a single
exponent, $H$ (the Hurst exponent) or by a composition of several sub-sets,
each one with a certain local exponent, $\alpha $, and all supported onto a main
structue. The former is defined as a \textit{mono-fractal} and the latter as a 
\textit{multi-fractal}. In this case the statistical proprieties of the
various sub-sets are characterised by the local exponents $\alpha $ are related with a
fractal dimension $f\left( \alpha \right) $\cite{feder} composing the \textit{%
multi-fractal spectrum}. To evaluate, numerically, this function we have
applied the MF-DFA method~\cite{mf-dfa}. For this procedure it was proved
that the $q$-th order fluctuation function, $F_{q}\left( s\right) $,
presents scale behaviour $F_{q}\left( s\right) \sim s^{h\left( q\right) }$.
The correspondence between MF-DFA and the standard formalism of
multifractals is obtained by, 
\begin{equation}
\tau \left( q\right) =q\,h\left( q\right) -1,
\end{equation}
where $\tau \left( q\right) $ is the exponent of the generalised partition
function. From Legendre transform, $f\left( \alpha \right) =q\,\alpha -\tau
\left( q\right) $, we can relate $\tau \left( q\right) $ with the H\"{o}lder 
exponent~\cite{feder}, $\alpha $. Thus, using the previous equation we get
\begin{equation}
\alpha =h\left( q\right) +q\frac{dh\left( q\right) }{dq},\qquad f\left(
\alpha \right) =q\,\left[ \alpha -h\left( q\right) \right] +1.
\end{equation}
In fig.~\ref{fig}(left) we display the $f\left( \alpha \right) $ spectrum (full line) obtained from 
averages for each $q$ over the values of the $30$ companies. In our analysis $q$ runs from $-20$ to $19.5$. 
We have verified that $f\left( \alpha \right) $ presents a wide range of
exponents from $\alpha _{\min }=0.32 \pm 0.04$ up to $\alpha _{\max }=1.09 \pm 0.04$,
corresponding to a deep multi-fractal behaviour. For $q=2$ we have obtained $%
h\left( 2\right) \equiv H=0.71 \pm 0.03$ which agrees with strong persistence
previously observed~\cite{volume}.
From our time series we can define new and related ones that can help
us to quantify which factors contribute the most to the multi-fractal
character of this observable. Among these factors we name: linear
correlations, non-linear correlations and power-law-like PDF. To that, we
have shuffled the elements (within each time series) and from
these series we have computed $h_{shuf}\left( q\right) $. From these
uncorrelated time series we have created another set by
randomising the phase in the Fourier space. Afterwards, we have applied the inverse Fourier transform 
to come back to the time variable. These new series have Gaussian stationary distribution
and scaling exponent $h_{sh-ran}\left( q\right) $. 

In fig.~\ref{fig}(left) we see that these two series also present multi-fractal 
spectrum, although the shuffle series has a wider spectrum than the shuffled plus phase randomised series. 
Concerning the Hurst exponent, $h(2)=H$, we have obtained $H=0.49 \pm 0.03$ for shuffled and
$H=0.5 \pm 0.03$ for shuffled plus phase randomised series. Considering error margins, these 
values are compatible with $H=1/2$ of Brownian motion. Furthermore, we have made a set of only
phase randomised time series for which we have obtained $H=0.7 \pm 0.03$. From this 
values we have concluded that correlations have a key role in the persistent character 
of traded volume time series.

Using $F_{q}\left(s\right) $ scaling relation for the three series \cite{mf-dfa} and assuming
that all the factors are independent, we can quantify the influence of
correlations, $h_{cor}\left( q\right) =h\left( q\right) -h_{shuf}\left(
q\right) $, the influence of a non-Gaussian PDF, $h_{PDF}\left( q\right)
=h_{shuf}\left( q\right) -h_{sh-ran}\left( q\right) $, and the weight of
non-linearities, $h_{sh-ran}\left( q\right) \equiv h_{nlin}\left( q\right)
=h\left( q\right) -h_{cor}\left( q\right) -h_{PDF}\left( q\right) $. The
multi-fractality of a time series can be analysed by mens of the difference of
values between $h\left( q_{\max }\right) $ and $h\left( q_{\min }\right) $,
hence 
\begin{equation}
\Delta h=h\left( q_{\min }\right) -h\left( q_{\max }\right)
\end{equation}
it is a suitable way to characterise multi-fractality. For a mono-fractal we
have $\Delta h=0$, \textit{i.e.}, linear dependence of $\tau \left( q\right) 
$ with $q$. In fig.~\ref{fig}(right)\ we have depicted $\tau \left( q\right) 
$ for several time series from which we have computed the various $\Delta h$%
. The results obtained are the following: $\Delta h=0.675$, $\Delta
h_{cor}=0.027$, $\Delta h_{PDF}=0.445$, and $\Delta h_{nlin}=0.203$. As it
can be easily concluded the influence of linear correlations in traded
volume muli-fractal nature is minimal with $\Delta h_{cor}$ corresponding to 
$4\%$ of $\Delta h$. This value is substancially smaller than the influence of $%
\Delta h_{nlin}$ which corresponds to $30\%$ of $\Delta h$. Our result is in
perfect accordance with another previous result of us \cite{vol-dumas}
where, using a non-extensive generalised mutual information measure~\cite{ct-kl}, 
we were able to show that non-linear dependences are not only stronger but
also more resilient than linear dependences (correlations) in traded volume time
series. Last but not least, from the values of $\Delta h$ we have verified
that the main factor for the multi-fractality of traded volume time series
is its non-Gaussian, generalised $q$-Gamma~\cite{volume,vol-smdq,vol-dumas}, 
probability density function with a weight of $66\%$ in $\Delta h$. 
Moreover, we have verified that the behaviour of $\tau \left( q\right) $ 
for $q>0$ is quite different from the $q<0$, which is also visible 
in the strong asymmetry of $f\left( \alpha \right) $. This could indicate 
that large and small fluctuations appear due to different dynamical 
mechanisms. Such scenario is consistent with the
super-statistical~\cite{beck-cohen} approach recently presented~\cite%
{vol-smdq,vol-dumas} and closely related with the current non-extensive
framework based on \textit{Tsallis entropy}~\cite{ct}. Within this context and 
bearing in mind the relation $1/(1-q_{sens})=1/\alpha_{min}-1/\alpha_{max}$~\cite{lyra-ct}, we
conjecture that, for traded volume, the sensivity to inicial conditions may be described by
$\xi = \left[ 1+\left( 1-q_{sens} \right) \, \lambda_{q_{sens}} \, t \right]^{\frac{1}{1-q_{sens}}}$ with
$q_{sens} = 0.55 \pm 0.08$.

\bigskip

We thank C.~Tsallis and E.~M.~F.~Curado for their continuous encouragement and valuable
discussions with P.~Ch.~Ivanov about phase randomisation. The data used in this work was provided by
Olsen Data Services to which we also ackowledge. This work benefitted from
infrastructural support from PRONEX (Brazilian agency) and financial support
of FCT/MCES (Portuguese agency).

\end{document}